\documentclass{article}
\usepackage{amssymb}
\usepackage{amsmath}

\input{tcilatex}

\begin{document}

\begin{center}
\textbf{Opportunities for use of exact statistical equations}

\textbf{Reginald J. Hill}

Cooperative Institute for Research in Environmental Sciences, University of
Colorado and National Oceanic and Atmospheric Administration, Environmental
Technology Laboratory 325 Broadway, Boulder CO, 80305-3328, United States of
America
\end{center}

\textbf{Abstract.} \ Exact structure function equations are an efficient
means of obtaining asymptotic laws such as inertial range laws, as well as
all measurable effects of inhomogeneity and anisotropy that cause deviations
from such laws. \ \textquotedblleft Exact\textquotedblright\ means that the
equations are obtained from the Navier-Stokes equation or other hydrodynamic
equations without any approximation. \ A pragmatic definition of local
homogeneity lies within the exact equations because terms that explicitly
depend on the rate of change of measurement location appear within the exact
equations; an analogous statement is true for local stationarity. \ An exact
definition of averaging operations is required for the exact equations. \
Careful derivations of several inertial range laws have appeared in the
literature recently in the form of theorems. \ These theorems give the
relationships of the energy dissipation rate to the structure function of
acceleration increment multiplied by velocity increment and to both the
trace of and the components of the third-order velocity structure functions.
\ These laws are efficiently derived from the exact velocity structure
function equations. \ In some respects, the results obtained herein differ
from the previous theorems.\ \ The acceleration-velocity structure function
is useful for obtaining the energy dissipation rate in particle tracking
experiments provided that the effects of inhomogeneity are estimated by
means of displacing the measurement location.

\section{Introduction}

\qquad Energy dissipation rate $\varepsilon $\ is used for scaling the
turbulence acceleration statistics that are measured by particle tracking at
Cornell University\cite{Vothetal02}. \ The inertial range of the
second-order velocity structure function is measured in the Cornell
apparatus to determine $\varepsilon $\ using the empirical inertial-range
relationship that requires local isotropy\cite{Vothetal02}. \ In the Cornell
experiments the flow between counter rotating blades produces high Reynolds
numbers but inhomogeneous, anisotropic turbulence. \ A motivation for the
present study was to use the exact velocity structure-function equation\cite%
{hillsecondorder02} to relate $\varepsilon $ to measurable quantities
without making any assumptions about homogeneity or isotropy of the flow and
hence, to obtain a relationship that applies exactly to the need to
determine $\varepsilon $\ in the Cornell experiment. \ By \textquotedblleft
exact\textquotedblright\ we mean that the equations follow from the
Navier-Stokes equation and\ the incompressibility condition with no
additional approximations. \ Because particle tracking is used at Cornell
University to measure acceleration and velocity, it is natural to relate the
acceleration-velocity structure function (i.e., the structure function of
the product of acceleration and velocity increments) to both $\varepsilon $\
and the third-order velocity structure function. \ On the basis of local
homogeneity, but without use of local isotropy, Mann, Ott and Andersen\cite%
{MannOttAnderson99,OttMann00} were the first to obtain an inertial-range
relationship of the acceleration-velocity structure function to $\varepsilon 
$. \ By specializing our exact equation for the acceleration-velocity
structure function to the locally homogeneous case, we efficiently obtain
their inertial-range result\ in Sec. 4, but an extraneous derivative moment
in their relationship is removed here. \ The exact structure function
equation method given here not only obtains that asymptotic law, but also
shows all quantities that must be evaluated to account for the effects of
inhomogeneity and anisotropy. \ Such evaluation requires the displacement of
the measurement volume that is defined by the Cornell laser beam and imaging
system.

\qquad The theorems of Nie and Tanveer\cite{NieTanveer99} and of Duchon and
Robert\cite{DuchonRobert00} establish the inertial-range 4/3 law that
relates $\varepsilon $\ to\ the trace of the third-order velocity structure
function. \ Instead of invoking local homogeneity, Nie and Tanveer\cite%
{NieTanveer99} perform a large volume average, and perform a long time
average instead of invoking local stationarity,\ and they use an average
over orientations of the two measurement points instead of invoking local
isotropy. \ Results\ of Nie and Tanveer\cite{NieTanveer99} are efficiently
obtained in Sec. 5 by use of the exact structure function equation. \ For
the case of spatially periodic direct numerical simulation, similar results
were obtained from the exact structure function equations in \cite%
{hillsecondorder02}.

\qquad The theorem of Duchon and Robert\cite{DuchonRobert00} differs from
that of Nie and Tanveer\cite{NieTanveer99} in an essential manner. \ The
space-time averaging required by Duchon and Robert is of arbitrary extent
and the viscosity is zero. \ In Sec. 6, using space-time averaging and the
limit of very large Reynolds number, the 4/3 law is obtained from the exact
statistical equations. \ The present derivation requires conditions on the
space-time average as stated in Sec. 6\ \ All terms that describe the
inhomogeneity and anisotropy of flows are retained. \ That result shows all
quantities that should be evaluated in direct numerical simulation (DNS) of
flows to determine causes of deviation from the 4/3 law.

\qquad The inertial-range 4/3 relationship of the trace of the third-order
velocity structure function is easily obtained relative to the
inertial-range 4/5 and 4/15 laws for the longitudinal and transverse
components of the third-order velocity structure function, respectively. \
It is remarkable that the 4/5 and 4/15 laws were obtained by Nie and Tanveer%
\cite{NieTanveer99} on the same basis as described above, and remarkable
that the 4/5 and 4/15 laws were also proven by Eyink\cite{Eyink03} on a
basis similar to that of Duchon and Robert\cite{DuchonRobert00}, that is,
for space-time averaging of arbitrary extent in the limit of vanishing
viscosity. \ Those 4/5 and 4/15 laws are not derived here. \ Eyink\cite%
{Eyink03} finds that tests of the inertial-range power laws for arbitrary
extent of averaging would be difficult. \ On the other hand, evaluation of
all terms in an exact statistical equation by means of DNS data would show
causes of deviations from those laws as well as the approach toward those
laws.

\qquad Of course, real experiments do not provide the opportunity to perform
the above mentioned averaging used in the derivation of the theorems for the
inertial-range laws. \ Deviations from 4/5 and 4/15 laws are observed
because of turbulence inhomogeneity, anisotropy, and finite Reynolds number.
\ Recently, there has been much work \cite%
{Lindborg99,Zhouetal00,Danaila01,Danaila02,Danaila04,DanailaNJP04} on
quantifying terms in the velocity structure function equations to learn how
the various terms affect the balance of the equation as a function of scale
and at what scales the effects of inhomogeneity and anisotropy become
pronounced and how asymptotic laws are approached. \ The exact structure
function equations are useful in this regard because they retain all effects
of inhomogeneity and anisotropy in a clearly organized manner. \ The exact
equations obviate the need for many derivations that obtain some limited
aspect of inhomogeneity. \ Those developments are discussed in Sec. 7. \ The
definition of local homogeneity that directly simplifies exact structure
function equations is that the rate of change of statistics with respect to
where they are measured is negligible. \ That pragmatic definition is
explained and contrasted with previous definitions in Sec. 2.

\section{Pragmatic definition of local homogeneity}

\qquad Exact statistical hydrodynamics involves the derivation of equations
relating statistics without the use of approximations. \ The Navier-Stokes
equation has been used to derive exact equations relating velocity structure
functions of velocity increments and other statistics\cite{hillsecondorder02}%
. \ Such exact equations have been obtained for all orders of velocity
structure functions in \cite{Hillallorderseqn01}. \ Structure function
equations of all orders have also been given for the asymptotic case of
local isotropy in \cite{Hillallorderseqn01}\ and \cite{Yakhot01}. \ The
exact equation for the scalar structure function has been obtained from the
continuity equation as well\cite{Hillscalar02}. \ All such exact equations
have optimal organization when the independent variables are chosen to be
the spatial separation\ $\mathbf{r}\equiv \mathbf{x}-\mathbf{x}^{\prime }$\
of the two points of measurement, i.e., $\mathbf{x}$ and $\mathbf{x}^{\prime
}$, and the midpoint $\mathbf{X}\equiv \left( \mathbf{x}+\mathbf{x}^{\prime
}\right) /2$. \ The derivative with respect to $\mathbf{X}$, i.e., $\partial
_{X_{n}}$, acts on some statistics within the structure function equations.
\ Local homogeneity is the approximation that the rate of change of
statistics with respect to the location of measurement may be neglected. \
Since that location is $\mathbf{X}$, the result of $\partial _{X_{n}}$\
acting on a statistic is neglected relative to other terms in the structure
function equations. \ This is a truly local definition; it makes no mention
of a spatial domain. \ This definition of local homogeneity was exploited in
the formulation of exact structure function equations in \cite%
{hillsecondorder02,Hillallorderseqn01}, and has been used in studies \cite%
{Danaila01,Danaila02,Danaila04,DanailaNJP04} of the effects of inhomogeneity
on the balance of structure function equations; it was introduced in \cite%
{Hill97}.

\qquad Local homogeneity has been given various definitions by different
authors. \ Kolmogorov\cite{Kolm41b} introduced a space-time domain that is
small compared to $L$ and $T=(L/U)$, where $L$ and $U$ are \textquotedblleft
typical length and velocity for the flow in the whole.\textquotedblright\ \
Kolmogorov considers the two-point differences of the velocities at spatial
points in the domain; one point is common to all the differences. \
Kolmogorov\cite{Kolm41b} defines local homogeneity as follows: \ the joint
probability distribution of the velocity differences is independent of the
one common spatial point, and of the velocity at the one common point, and
of time. \ There are data \cite%
{Praskovskyetal93,SreeniStolo96,SreeniDhruva98} that contradict the
statistical independence of velocity difference and the velocity at either
end point, and also contradict the statistical independence of velocity
difference and the velocity at the midpoint. \ The exception is isotropic
turbulence \cite{SreeniDhruva98},\ for which case local homogeneity is
assured. \ An alternative possibility that is particularly relevant here is
that the two-point velocity sum, $u_{n}+u_{n}^{\prime }$, might be
statistically independent of velocity difference, but statements in \cite%
{SreeniStolo96,SreeniDhruva98}\ contradict that statistical independence as
well. \ Therefore, Kolmogorov's\cite{Kolm41b} definition should not be used
because experimental data contradict that statistical independence \cite%
{Praskovskyetal93,SreeniStolo96,HillWilczak01}, as do theoretical
considerations \cite{HillWilczak01}. \ Recently, Frisch et al.\cite%
{Frischetal05} have considered the inconsistency of Kolmogorov's definition
of local homogeneity.

\qquad Monin and Yaglom \cite{MoninYaglom75} define local homogeneity to
mean that the joint probability distribution of the two-spatial-point
velocity differences is unaffected by any translation of the spatial points.
\ They do not impose a restriction on the translations to a spatial domain.
\ It follows that statistics composed entirely of the velocity differences
obey the same relationships that they do for homogeneous turbulence (namely,
they are independent of where they are measured), and that the mean velocity
depends linearly on position\cite{MoninYaglom75}. \ In practice, statistics
of velocity differences and of derivatives do depend on where they are
measured except in the ideal case of homogeneous turbulence. \ Frisch \cite%
{Frisch95} gives a definition that is equivalent to that of Monin and Yaglom 
\cite{MoninYaglom75}, except that the translations are restricted to a
domain the size of the spatial scale characteristic of the production of
turbulent energy (which he calls the \textquotedblleft integral
scale\textquotedblright ). \ Two-point structure function equations of all
orders contain a statistic that is the product of not only factors of the
velocity difference but also one factor of the sum of the two velocities,
i.e., $u_{n}+u_{n}^{\prime }$ \cite{hillsecondorder02,Hillallorderseqn01}. \
Because the definitions of local homogeneity by Monin and Yaglom \cite%
{MoninYaglom75} and Frisch \cite{Frisch95} involve only the joint
probability distribution of two-point differences, but do not involve $%
u_{n}+u_{n}^{\prime }$, it follows that those definitions are not sufficient
to simplify structure function equations to the same level of simplification
as does homogeneity.

\section{Definitions and notation}

\subsection{Definition of two-point, two-time quantities}

\qquad Here, we extend the results in \cite{hillsecondorder02} to two times
and to the structure function composed of the product of acceleration
difference and velocity difference. \ Four independent variables are
considered; $\mathbf{x}$ and $\mathbf{x}^{\prime }$ are spatial points; $t$\
and $t^{\prime }$\ are times. \ Denote velocities by $u_{i}=u_{i}(\mathbf{x}%
,t)$, $u_{i}^{\prime }=u_{i}(\mathbf{x}^{\prime },t^{\prime })$, and
accelerations by\ $a_{i}=a_{i}(\mathbf{x},t)$, $a_{i}^{\prime }=a_{i}(%
\mathbf{x}^{\prime },t^{\prime })$, etc. \ Quantities that will appear when
we use the Navier-Stokes equation are%
\begin{eqnarray*}
d_{ij} &\equiv &\left( u_{i}-u_{i}^{\prime }\right) \left(
u_{j}-u_{j}^{\prime }\right) \\
d_{ijn} &\equiv &\left( u_{i}-u_{i}^{\prime }\right) \left(
u_{j}-u_{j}^{\prime }\right) \left( u_{n}-u_{n}^{\prime }\right) \\
\digamma _{ijn} &\equiv &\left( u_{i}-u_{i}^{\prime }\right) \left(
u_{j}-u_{j}^{\prime }\right) \frac{u_{n}+u_{n}^{\prime }}{2} \\
\tau _{ij} &\equiv &\left( \partial _{x_{i}}p-\partial _{x_{i}^{\prime
}}p^{\prime }\right) \left( u_{j}-u_{j}^{\prime }\right) +\left( \partial
_{x_{j}}p-\partial _{x_{j}^{\prime }}p^{\prime }\right) \left(
u_{i}-u_{i}^{\prime }\right) \\
e_{ij} &\equiv &\left( \partial _{x_{n}}u_{i}\right) \left( \partial
_{x_{n}}u_{j}\right) +\left( \partial _{x_{n}^{\prime }}u_{i}^{\prime
}\right) \left( \partial _{x_{n}^{\prime }}u_{j}^{\prime }\right) \\
A_{ij} &\equiv &\left( a_{i}-a_{i}^{\prime }\right) \left(
u_{j}-u_{j}^{\prime }\right) +\left( a_{j}-a_{j}^{\prime }\right) \left(
u_{i}-u_{i}^{\prime }\right) ,
\end{eqnarray*}%
where $\partial _{x_{i}}p$ is pressure gradient. \ Change of variables:%
\begin{equation}
\mathbf{X}\equiv \left( \mathbf{x}+\mathbf{x}^{\prime }\right) /2\text{ \
and \ }\mathbf{r}\equiv \mathbf{x}-\mathbf{x}^{\prime }\text{; \ \c{T}}%
=\left( t+t^{\prime }\right) /2\text{ \ and \ \c{t}}\equiv t-t^{\prime }.
\label{variable change}
\end{equation}%
Just as $\left( \mathbf{x},\mathbf{x}^{\prime },t,t^{\prime }\right) $\ are
independent variables, so are $\left( \mathbf{X},\mathbf{r},\text{\c{T}},%
\text{\c{t}}\right) $. \ The significance of variable $\mathbf{X}$ is that
it is the location of measurement. \ Nonzero values of derivatives of
statistics with respect to $\mathbf{X}$, evaluated at position $\mathbf{X}$,
are the result of local inhomogeneity of the flow. \ Likewise, nonzero
values of derivatives of statistics with respect to \c{T} are the result of
nonstationarity of the flow. The relationships between the spatial
derivatives are%
\[
\partial _{x_{i}}=\partial _{r_{i}}+\frac{1}{2}\partial _{X_{i}}\text{%
{\small , \ }}\partial _{x_{i}^{\prime }}=-\partial _{r_{i}}+\frac{1}{2}%
\partial _{X_{i}}\text{, \ }\partial _{X_{i}}=\partial _{x_{i}}+\partial
_{x_{i}^{\prime }}\text{{\small , \ }}\partial _{r_{i}}=\frac{1}{2}\left(
\partial _{x_{i}}-\partial _{x_{i}^{\prime }}\right) ,
\]%
which give the useful properties%
\begin{equation}
\partial _{r_{i}}\left[ f(\mathbf{x},t)\pm g(\mathbf{x}^{\prime },t^{\prime
})\right] =\partial _{X_{i}}\left[ f(\mathbf{x},t)\mp g(\mathbf{x}^{\prime
},t^{\prime })\right] /2.  \label{r and X derivatives}
\end{equation}%
Similarly, 
\[
\partial _{\text{\c{T}}}=\partial _{t}+\partial _{t^{\prime }}\text{, \ }%
\partial _{\text{\c{t}}}\equiv \frac{1}{2}\left( \partial _{t}-\partial
_{t^{\prime }}\right) .
\]%
Use of the derivative formulas on $\tau _{ij}$ gives

\begin{equation}
\tau _{ij}=-2\left( p-p^{\prime }\right) \left( s_{ij}-s_{ij}^{\prime
}\right) +\partial _{X_{i}}\left[ \left( p-p^{\prime }\right) \left(
u_{j}-u_{j}^{\prime }\right) \right] +\partial _{X_{j}}\left[ \left(
p-p^{\prime }\right) \left( u_{i}-u_{i}^{\prime }\right) \right] ,
\label{taoij}
\end{equation}%
where\ \ $s_{ij}\equiv \left( \partial _{x_{i}}u_{j}+\partial
_{x_{j}}u_{i}\right) /2$ is the rate of strain. \ The trace of $\tau _{ij}$\
is%
\begin{equation}
\tau _{ii}=2\partial _{X_{i}}\left[ \left( p-p^{\prime }\right) \left(
u_{i}-u_{i}^{\prime }\right) \right] .  \label{tao trace}
\end{equation}%
Use of derivative formulas on $e_{ij}$ and taking the trace and use of
Poisson's equation, i.e., $\partial _{x_{n}}\partial _{x_{n}}p=-\partial
_{x_{i}}u_{j}\partial _{x_{j}}u_{i}$, gives%
\[
e_{ii}=\nu ^{-1}\left( \varepsilon +\varepsilon ^{\prime }\right) +\partial
_{X_{n}}\partial _{X_{n}}\left( p+p^{\prime }\right) ,
\]%
where\ $\varepsilon \equiv 2\nu s_{ij}s_{ij}$ is the energy dissipation rate.

\subsection{Use of the Navier-Stokes equation}

\qquad The Navier-Stokes equation at $(\mathbf{x},t)$\ is%
\[
a_{i}=\partial _{t}u_{i}+u_{n}\partial _{x_{n}}u_{i}=-\partial _{x_{i}}p+\nu
\partial _{x_{n}}\partial _{x_{n}}u_{i}\text{ , and\ }\partial
_{x_{n}}u_{n}=0.
\]%
Subtracting the Navier-Stokes equations at $(\mathbf{x}^{\prime },t^{\prime
})$, multiplying by $\left( u_{j}-u_{j}^{\prime }\right) $, and adding the
equation needed to produce symmetry under interchange of $i$\ and $j$\ gives%
\begin{equation}
A_{ij}=\partial _{\text{\c{T}}}d_{ij}+\partial _{X_{n}}\digamma
_{ijn}+\partial _{r_{n}}d_{ijn}=-\tau _{ij}+2\nu \left( \partial
_{r_{n}}\partial _{r_{n}}d_{ij}+\frac{1}{4}\partial _{X_{n}}\partial
_{X_{n}}d_{ij}-e_{ij}\right) .  \label{basic ij eqn}
\end{equation}%
The trace gives 
\begin{equation}
A_{ii}=\partial _{\text{\c{T}}}d_{ii}+\partial _{X_{n}}\digamma
_{iin}+\partial _{r_{n}}d_{iin}=2\nu \partial _{r_{n}}\partial
_{r_{n}}d_{ii}-2\left( \varepsilon +\varepsilon ^{\prime }\right) +W,
\label{trace}
\end{equation}%
where 
\begin{equation}
W\equiv -2\partial _{X_{i}}\left[ \left( p-p^{\prime }\right) \left(
u_{i}-u_{i}^{\prime }\right) \right] +\frac{\nu }{2}\partial
_{X_{n}}\partial _{X_{n}}d_{ii}-2\nu \partial _{X_{n}}\partial
_{X_{n}}\left( p+p^{\prime }\right) .  \label{Wtrace}
\end{equation}%
No average has been used yet.

\subsection{The $\mathbf{r}$-sphere\ and orientation averages}

\qquad Local isotropy has been used in the past to remove the divergence
from $\partial _{r_{n}}d_{iin}$ and to proceed to Kolmogorov's equation. \
Without the assumption of local isotropy, an average over an $\mathbf{r}$%
-sphere removes the divergence from $\partial _{r_{n}}d_{iin}$. \ The $%
\mathbf{r}$-sphere\ average also mitigates anisotropy\cite%
{TaylorKurienEyink03,hillsecondorder02}. \ Energy dissipation rate $%
\varepsilon $ averaged over a sphere in $\mathbf{r}$-space, $\left\langle
\varepsilon \right\rangle _{\mathbf{r}\text{-sphere}}$, was introduced by
Obukhov\cite{Obukhov62} and Kolmogorov\cite{Kolmogorov62} in 1962; it is a
recurrent theme in small-scale similarity theories. \ We have produced exact
dynamical equations containing the sphere-averaged energy dissipation rate%
\cite{hillsecondorder02}. \ The volume average over an $\mathbf{r}$-space
sphere of radius $r_{S}$\ of a quantity $Q$\ is defined by 
\begin{equation}
\left\langle Q\right\rangle _{\mathbf{r}\text{-sphere}}\text{ }\left( 
\mathbf{X,}r_{S},\text{\c{T}},\text{\c{t}}\right) \equiv \left( 4\pi
r_{S}^{3}/3\right) ^{-1}\underset{\left\vert \mathbf{r}\right\vert \text{ }%
\leq \text{ }r_{S}}{\int \int \int }Q\left( \mathbf{X,r,}\text{\c{T}},\text{%
\c{t}}\right) d\mathbf{r.}  \label{r sphere average}
\end{equation}%
The orientation average over the surface of the $\mathbf{r}$-space sphere of
radius $r_{S}$ of the outward normal component of any vector $q_{n}$ is
defined by%
\begin{eqnarray}
\oint_{\mathbf{r}_{n}}q_{n} &\equiv &\left( 4\pi r_{S}^{2}\right) ^{-1}%
\underset{\left\vert \mathbf{r}\right\vert \text{ }=\text{ }r_{S}}{\int \int 
}\frac{r_{n}}{r}q_{n}\left( \mathbf{X,r},t\right) ds
\label{orientation average} \\
&=&\left( 4\pi \right) ^{-1}\underset{\left\vert \mathbf{r}\right\vert \text{
}=\text{ }r_{S}}{\int \int }\frac{r_{n}}{r}q_{n}\left( \mathbf{X,r},t\right)
d\Omega ,  \nonumber
\end{eqnarray}%
where $ds$\ is the differential of surface area and $d\Omega $\ is the
differential of solid angle. \ In (\ref{orientation average}), the inner
product $\frac{r_{n}}{r}q_{n}$\ produces the so-called longitudinal
component of $q_{n}$. \ The above averages can be performed on data. \ The
divergence theorem is 
\begin{equation}
\left\langle \partial _{r_{n}}q_{n}\right\rangle _{\mathbf{r}\text{-sphere}%
}=\left( 3/r_{S}\right) \oint_{\mathbf{r}_{n}}q_{n}.
\label{divergence theorem r space}
\end{equation}

\subsection{The $\mathbf{X}$-space and $\mathbf{X}$-surface averages}

\qquad Let the spatial average be over a region $\mathbb{R}$ in $\mathbf{X}$%
-space. \ The spatial average of any quantity $Q$ is denoted by $%
\left\langle Q\right\rangle _{\mathbb{R}}$, which has argument list $\left( 
\mathbb{R}\mathbf{,r},\text{\c{T}},\text{\c{t}}\right) $; that is, 
\[
\left\langle Q\right\rangle _{\mathbb{R}}\text{ }\left( \mathbb{R}\mathbf{,r}%
,\text{\c{T}},\text{\c{t}}\right) \equiv \frac{1}{V}\underset{\mathbb{R}}{%
\int \int \int }Q\left( \mathbf{X,r},\text{\c{T}},\text{\c{t}}\right) d%
\mathbf{X,}
\]%
where $V$ is the volume of the space region $\mathbb{R}$. \ Given any vector 
$q_{n}$, the divergence theorem relates the volume average of $\partial
_{X_{n}}q_{n}$ to the surface average; that is, 
\begin{equation}
\left\langle \partial _{X_{n}}q_{n}\right\rangle _{\mathbb{R}}\equiv \frac{1%
}{V}\int \int \int \partial _{X_{n}}q_{n}d\mathbf{X}=\frac{S}{V}\left( \frac{%
1}{S}\int \int \widehat{N}_{n}q_{n}dS\right) \equiv \frac{S}{V}\oint_{%
\mathbf{X}_{n}}q_{n},  \label{Xsurfave}
\end{equation}%
where $S$ is the surface area bounding $\mathbb{R}$, $dS$ is the
differential of surface area, and $\widehat{N}_{n}$ is the unit vector
oriented outward and normal to the surface. \ As seen on the right-hand side
of (\ref{Xsurfave}), we adopt, for brevity, the integral-operator notation 
\[
\oint_{\mathbf{X}_{n}}\equiv \frac{1}{S}\int \int \widehat{N}_{n}dS.
\]

\subsection{Time average}

\qquad Consider the term $\partial _{\text{\c{T}}}d_{ij}$ in (\ref{basic ij
eqn}). \ The time average of a quantity $Q$\ from an initial time \c{T}$_{0}$%
\ and over a duration $T$\ is defined by%
\begin{equation}
\left\langle Q\right\rangle _{T}\text{ }\left( \mathbf{X,r},\text{\c{T}}%
_{0},T,\text{\c{t}}\right) \equiv \frac{1}{T}\int_{\text{\c{T}}_{0}}^{\text{%
\c{T}}_{0}+T}Q\left( \mathbf{X,r},\text{\c{T}},\text{\c{t}}\right) \text{ }d%
\text{\c{T}}.  \label{timeavedef}
\end{equation}%
Of course, $\partial _{\text{\c{T}}}$\ does not commute with the integral
operator (\ref{timeavedef}); it follows that 
\begin{eqnarray}
\left\langle \partial _{\text{\c{T}}}d_{ij}\right\rangle _{T} &\equiv &\frac{%
1}{T}\int_{\text{\c{T}}_{0}}^{\text{\c{T}}_{0}+T}\left( \partial _{\text{%
\c{T}}}d_{ij}\right) \text{ }d\text{\c{T}}  \nonumber \\
&=&\left[ d_{ij}\left( \mathbf{X,r},\text{\c{T}}_{0}+T,\text{\c{t}}\right)
-d_{ij}\left( \mathbf{X,r},\text{\c{T}}_{0},\text{\c{t}}\right) \right] /T.
\label{ave time deriv}
\end{eqnarray}%
This shows that it is easy to evaluate $\left\langle \partial _{\text{\c{T}}%
}d_{ij}\right\rangle _{T}$\ using experimental data because only the data at
times \c{T}$_{0}$\ and \c{T}$_{0}+T$ are used. \ One can make $\left\langle
\partial _{\text{\c{T}}}d_{ij}\right\rangle _{T}$ as small as one desires by
allowing $T$ to be very large, provided that $d_{ij}\left( \mathbf{X,r},%
\text{\c{T}}_{0}+T,\text{\c{t}}\right) $ does not differ greatly from $%
d_{ij}\left( \mathbf{X,r},\text{\c{T}}_{0},\text{\c{t}}\right) $.

\subsection{All averages commute}

\qquad The above averages are integrations with respect to independent
variables\ $\mathbf{X}$, $\mathbf{r}$, and \c{T}; also, an ensemble average
is a sum over realizations. \ All those averages commute with one another. \
They commute with derivatives with the following exceptions: \ Volume and
surface averages $\left\langle Q\right\rangle _{\mathbb{R}}$\ and $\oint_{%
\mathbf{X}_{n}}$\ do not commute with the derivative\textbf{\ }$\partial
_{X_{i}}$, nor do $\left\langle Q\right\rangle _{\mathbf{r}\text{-sphere}}$\
and $\oint_{\mathbf{r}_{n}}q_{n}$\ commute with\textbf{\ }$\partial _{r_{i}}$%
, nor does $\left\langle Q\right\rangle _{T}$\ commute with $\partial _{%
\text{\c{T}}}$.

\section{Theorem of Mann, Ott, and Andersen}

\qquad Mann et al.\cite{MannOttAnderson99} and Ott and Mann\cite{OttMann00}
studied turbulent dispersion of particles by means of particle tracking. \
They therefore used an ensemble average over particle trajectory events. \
The ensemble average is formed from observations of many particle-pair
trajectories and $\mathbf{r}$\textbf{\ }values. \ It is also possible to use
single trajectories; then two times, $t$ and $t^{\prime }$, are needed. \
From the data, averages are stored in bins of the direction and length of $%
\mathbf{r}$. \ Denote the ensemble average by $\left\langle {}\right\rangle
_{E}$. \ The ensemble average is a sum and therefore commutes with the
spatial and temporal derivatives. \ Mann et al.\cite{MannOttAnderson99}
assumed local homogeneity. \ From the present definition of local
homogeneity, we have $\left\langle \partial _{X_{n}}\digamma
_{iin}\right\rangle _{E}=\partial _{X_{n}}\left\langle \digamma
_{iin}\right\rangle _{E}=0$, and similarly $\left\langle W\right\rangle
_{E}=0$. \ Consequently, the ensemble average of (\ref{trace}) can be
written as%
\begin{equation}
\left\langle A_{ii}\right\rangle _{E}=\partial _{\text{\c{T}}}\left\langle
d_{ii}\right\rangle _{E}+\partial _{r_{n}}\left\langle d_{iin}\right\rangle
_{E}=2\nu \partial _{r_{n}}\partial _{r_{n}}\left\langle d_{ii}\right\rangle
_{E}-2\left\langle \varepsilon +\varepsilon ^{\prime }\right\rangle _{E}.
\label{MannOtt first}
\end{equation}%
For $\mathbf{r}$\ much larger than dissipation-range scales, Mann et al.\cite%
{MannOttAnderson99} neglect the term $2\nu \partial _{r_{n}}\partial
_{r_{n}}\left\langle d_{ii}\right\rangle _{E}$. \ Doing so, and using the
definition of $A_{ij}$, we have%
\begin{equation}
\left\langle \left( a_{i}-a_{i}^{\prime }\right) \left( u_{i}-u_{i}^{\prime
}\right) \right\rangle _{E}=-\left\langle \varepsilon +\varepsilon ^{\prime
}\right\rangle _{E}.  \label{MannOttIR}
\end{equation}%
With two exceptions, this is the result of \cite{MannOttAnderson99} in their
equation (90) \ First, their result contains an extraneous derivative moment
that is absent above. \ Second, $\left\langle \varepsilon +\varepsilon
^{\prime }\right\rangle _{E}$\ appears above, whereas $2\left\langle
\varepsilon \right\rangle _{E}$\ is in their result.\ \ In an experimental
apparatus, there can be a systematic difference between the energy
dissipation rates at $(\mathbf{x},t)$\ versus $(\mathbf{x}^{\prime
},t^{\prime })$, even if $t=t^{\prime }$, such that $\left\langle
\varepsilon \right\rangle _{E}\neq \left\langle \varepsilon ^{\prime
}\right\rangle _{E}$.

\qquad The above ensemble averages depend on $\mathbf{r}$, which is the
vector separation of points on trajectories from which data are obtained. \
That dependence on $\mathbf{r}$ allows a further average over the $\mathbf{r}
$-sphere. \ Anisotropy of the flow might favor the occurrence of some
orientations of $\mathbf{r}$; by weighting of the occurrences, the $\mathbf{r%
}$-sphere average is constructed for uniform orientation of $\mathbf{r}$.

\subsection{Applying the $\mathbf{r}$-sphere\ average}

\qquad The $\mathbf{r}$-sphere average causes each point $\mathbf{x}$\ to
coincide with point $\mathbf{x}^{\prime }$\ when the orientation of $\mathbf{%
r}$\ is reversed. \ Consequently, 
\[
\left\langle \left\langle \varepsilon +\varepsilon ^{\prime }\right\rangle
_{E}\right\rangle _{\mathbf{r}\text{-sphere}}=2\left\langle \left\langle
\varepsilon \right\rangle _{E}\right\rangle _{\mathbf{x}\text{-sphere}%
}=2\left\langle \left\langle \varepsilon ^{\prime }\right\rangle
_{E}\right\rangle _{\mathbf{x}^{\prime }\text{-sphere}}
\]%
where averages over spheres in $\mathbf{x}$-space and $\mathbf{x}^{\prime }$%
-space are introduced. \ The particle tracking experiment is temporally
stationary such that $\partial _{\text{\c{T}}}\left\langle
d_{ii}\right\rangle _{E}$\ vanishes. \ Then, the $\mathbf{r}$-sphere\
average of (\ref{MannOtt first}) gives%
\[
\oint_{\mathbf{r}_{n}}\left\langle d_{iin}\right\rangle _{E}=-\frac{4}{3}%
r_{S}\left\langle \left\langle \varepsilon \right\rangle _{E}\right\rangle _{%
\mathbf{x}\text{-sphere}},
\]%
which is an extension of the 4/3 law without use of local isotropy, and

\begin{equation}
\left\langle \left\langle \left( a_{i}-a_{i}^{\prime }\right) \left(
u_{i}-u_{i}^{\prime }\right) \right\rangle _{E}\right\rangle _{\mathbf{r}%
\text{-sphere}}=-2\left\langle \left\langle \varepsilon \right\rangle
_{E}\right\rangle _{\mathbf{x}\text{-sphere}}.  \label{MannOttsecond}
\end{equation}%
Mann et al.\cite{MannOttAnderson99} give (\ref{MannOttsecond}), except that
their result contains an extraneous derivative moment. \ See equation (91)
in \cite{MannOttAnderson99}.

\subsection{Quantifying effects of inhomogeneity}

\qquad The terms that are neglected on the basis of local homogeneity, such
as $\partial _{X_{n}}\left\langle \digamma _{iin}\right\rangle _{E}$, can be
measured by particle tracking. \ To evaluate the effects of inhomogeneity,
the rate of change with respect to displacement of the averaging volume is
needed, i.e., $\mathbf{X}$\ must be changed. \ The most troublesome of the
terms describing inhomogeneity is the term $-2\partial _{X_{i}}\left[ \left(
p-p^{\prime }\right) \left( u_{i}-u_{i}^{\prime }\right) \right] $ in (\ref%
{Wtrace}) because it requires a measurement of pressure difference. \ By
applying the derivative relationship (\ref{r and X derivatives}), the $%
\mathbf{r}$-sphere\ average (\ref{r sphere average}), and the divergence
theorem (\ref{divergence theorem r space}), the most troublesome term can be
expressed as%
\[
\partial _{X_{i}}\left\langle \left\langle \left( p-p^{\prime }\right)
\left( u_{i}-u_{i}^{\prime }\right) \right\rangle _{E}\right\rangle _{%
\mathbf{r}\text{-sphere}}=\frac{6}{r_{S}}\oint_{\mathbf{r}_{n}}\left\langle
\left( p+p^{\prime }\right) \left( u_{n}-u_{n}^{\prime }\right)
\right\rangle _{E}.
\]%
The right-hand side requires the two-point pressure sum correlated with the
velocity difference and an average over orientations of $\mathbf{r}$ as
defined in(\ref{orientation average}).

\section{Theorem of Nie and Tanveer}

\qquad The spatial average of (\ref{trace}) and use of the divergence
theorem (\ref{Xsurfave}) gives 
\begin{eqnarray}
\left\langle A_{ii}\right\rangle _{\mathbb{R}} &=&\partial _{\text{\c{T}}%
}\left\langle d_{ii}\right\rangle _{\mathbb{R}}+\frac{S}{V}\oint_{\mathbf{X}%
_{n}}\digamma _{iin}+\partial _{r_{n}}\left\langle d_{iin}\right\rangle _{%
\mathbb{R}}  \nonumber \\
&=&2\nu \partial _{r_{n}}\partial _{r_{n}}\left\langle d_{ii}\right\rangle _{%
\mathbb{R}}-2\left\langle \varepsilon +\varepsilon ^{\prime }\right\rangle _{%
\mathbb{R}}+\left\langle W\right\rangle _{\mathbb{R}},
\label{space ave trace}
\end{eqnarray}%
where%
\begin{equation}
\left\langle W\right\rangle _{\mathbb{R}}\equiv \frac{S}{V}\oint_{\mathbf{X}%
_{n}}\left[ -2\left( p-p^{\prime }\right) \left( u_{n}-u_{n}^{\prime
}\right) +\frac{\nu }{2}\partial _{X_{n}}d_{ij}-2\nu \partial _{X_{n}}\left(
p+p^{\prime }\right) \right] .  \label{W space averaged}
\end{equation}%
These averages are for any arbitrary region $\mathbb{R}$.

\qquad There are several possibilities for eliminating the terms $\frac{S}{V}%
\oint_{\mathbf{X}_{n}}\digamma _{iin}$ and $\left\langle W\right\rangle _{%
\mathbb{R}}$. \ First, those terms describe the effects of inhomogeneity and
are negligible in the case of local homogeneity, which case requires a
sufficiently large Reynolds number and small $\left\vert \mathbf{r}%
\right\vert $. \ Second, as in \cite{hillsecondorder02}, those terms vanish
for the case of periodic boundary conditions, such as are often used in DNS,
provided that the spatial average is over the entire spatial period. \
Third, those terms vanish if the turbulence is of limited spatial extent and
the volume average is over a much larger region such that the velocity and
pressure fields are negligible on the surface that bounds $\mathbb{R}$; this
possibility causes the statistics to decrease as $V$\ increases. \ Fourth,
if the region $\mathbb{R}$\ is enclosed by rigid boundaries with the no-slip
boundary condition on velocities, then only the pressure gradient term at
far right in (\ref{W space averaged}) is nonzero at points on the
boundaries; a separate hypothesis that $\oint_{\mathbf{X}_{n}}\partial
_{X_{n}}\left( p+p^{\prime }\right) $\ vanishes is required. \ Nie and
Tanveer\cite{NieTanveer99} eliminate the subject terms by integrating over
the \textquotedblleft entire volume\textquotedblright ;\ the meaning must be
of an infinite volume unless boundary conditions are specified. \ Neglecting
the subject terms, we have%
\begin{equation}
\left\langle A_{ii}\right\rangle _{\mathbb{R}}=\partial _{\text{\c{T}}%
}\left\langle d_{ii}\right\rangle _{\mathbb{R}}+\partial
_{r_{n}}\left\langle d_{iin}\right\rangle _{\mathbb{R}}=2\nu \partial
_{r_{n}}\partial _{r_{n}}\left\langle d_{ii}\right\rangle _{\mathbb{R}%
}-2\left\langle \varepsilon +\varepsilon ^{\prime }\right\rangle _{\mathbb{R}%
}.  \label{pre Duchon}
\end{equation}%
For $\mathbf{r}$\ much larger than dissipation-range scales, neglect the
term $2\nu \partial _{r_{n}}\partial _{r_{n}}\left\langle
d_{ii}\right\rangle _{\mathbb{R}}$. \ For an average over the entire flow
and for $t=t^{\prime }$, 
\[
\left\langle \varepsilon +\varepsilon ^{\prime }\right\rangle _{\mathbb{R}%
}=2\left\langle \varepsilon \right\rangle _{\mathbb{R}}=2\left\langle
\varepsilon ^{\prime }\right\rangle _{\mathbb{R}}.
\]%
If $t\neq t^{\prime }$, then clearly, $\left\langle \varepsilon
\right\rangle _{\mathbb{R}}\neq \left\langle \varepsilon ^{\prime
}\right\rangle _{\mathbb{R}}$. \ The definition of $A_{ij}$\ and (\ref{pre
Duchon}) give%
\begin{equation}
\left\langle \left( a_{i}-a_{i}^{\prime }\right) \left( u_{i}-u_{i}^{\prime
}\right) \right\rangle _{\mathbb{R}}=-2\left\langle \varepsilon
\right\rangle _{\mathbb{R}}.  \label{NieTanveer accelvelo}
\end{equation}

\qquad Apply the time average (\ref{timeavedef}) to (\ref{pre Duchon}). \
Nie and Tanveer\cite{NieTanveer99} take the averaging time, $T$, to be
infinite such that (\ref{ave time deriv}) vanishes. \ Alternatively, one can
assume sufficient stationarity that the trace of (\ref{ave time deriv}) is
negligible. \ Also apply the $\mathbf{r}$-sphere\ average (\ref{r sphere
average}). \ Then (\ref{pre Duchon}) gives, for $t=t^{\prime }$,%
\begin{equation}
\oint_{\mathbf{r}_{n}}\left\langle \left\langle d_{iin}\right\rangle _{%
\mathbb{R}}\right\rangle _{T}=-\frac{4}{3}r_{S}\left\langle \left\langle
\left\langle \varepsilon \right\rangle _{\mathbb{R}}\right\rangle
_{T}\right\rangle _{\mathbf{r}\text{-sphere}}.  \label{Duychon and Rob}
\end{equation}%
Note that the orientation average of $d_{iin}$,\ as defined in (\ref%
{orientation average}), appears in (\ref{Duychon and Rob}) as a consequence
of the divergence theorem (\ref{divergence theorem r space}). \ This is a
theorem of Nie and Tanveer\cite{NieTanveer99}. \ Their order of averaging is
different, namely $\mathbb{R}$\ then $\oint_{\mathbf{r}_{n}}$\textbf{\ }then 
$T$, but the averages commute as discussed in Sec. 3.6.

\section{Theorem of Duchon and Robert}

\qquad The important distinction between the theorem of Nie and Tanveer
(discussed above) and that of Duchon and Robert\cite{DuchonRobert00} is that
the space-time averaging required by Duchon and Robert is of arbitrary
extent and the viscosity is zero. \ In this case, the averaging region $%
\mathbb{R}$\ is a subdomain of the entire flow. \ Recall that $d_{ij}\equiv
\left( u_{i}-u_{i}^{\prime }\right) \left( u_{j}-u_{j}^{\prime }\right) $,
and write 
\begin{equation}
\digamma _{ijn}=\left\langle \breve{u}_{n}\right\rangle _{\mathbb{R}}d_{ij}+%
\frac{\widehat{u}_{n}+\widehat{u}_{n}^{\prime }}{2}d_{ij},
\label{F decomposition}
\end{equation}%
where%
\[
\widehat{u}_{n}\equiv u_{n}-\left\langle u_{n}\right\rangle _{\mathbb{R}}%
\text{, \ }\widehat{u}_{n}^{\prime }\equiv u_{n}^{\prime }-\left\langle
u_{n}^{\prime }\right\rangle _{\mathbb{R}}\text{, \ and \textbf{\ }}\breve{u}%
_{n}\equiv \frac{u_{n}+u_{n}^{\prime }}{2}
\]%
Recall that (\ref{space ave trace})-(\ref{W space averaged})\ apply for any
arbitrary space averaging region $\mathbb{R}$. \ Let $\nu =0$\ in (\ref%
{space ave trace})-(\ref{W space averaged}), substitute (\ref{F
decomposition}), and multiply by $r_{S}/3$, then%
\begin{eqnarray}
&&\frac{r_{S}}{3}\left\langle A_{ii}\right\rangle _{\mathbb{R}}=\frac{r_{S}}{%
3}\partial _{\text{\c{T}}}\left\langle d_{ii}\right\rangle _{\mathbb{R}}+%
\frac{r_{S}}{3}\left\langle \left\langle \breve{u}_{n}\right\rangle _{%
\mathbb{R}}\partial _{X_{n}}d_{ii}\right\rangle _{\mathbb{R}}  \nonumber \\
&&+\frac{1}{3}\frac{Sr_{S}}{V}\oint_{\mathbf{X}_{n}}\frac{\widehat{u}_{n}+%
\widehat{u}_{n}^{\prime }}{2}d_{ij}+\frac{r_{S}}{3}\partial
_{r_{n}}\left\langle d_{iin}\right\rangle _{\mathbb{R}}  \nonumber \\
&=&-\frac{4}{3}r_{S}\left\langle \frac{\varepsilon +\varepsilon ^{\prime }}{2%
}\right\rangle _{\mathbb{R}}-\frac{2}{3}\frac{Sr_{S}}{V}\oint_{\mathbf{X}%
_{n}}\left( p-p^{\prime }\right) \left( u_{n}-u_{n}^{\prime }\right) .
\label{TRD1}
\end{eqnarray}%
Take the averaging region $\mathbb{R}$\ to have a simple topology such that
the volume to surface ratio $V/S$\ is the size of $\mathbb{R}$.\textbf{\ } \
Because $\nu =0$\ is effectively the limit of infinite Reynolds number, $%
r_{S}$\ can be as small as desired. \ Therefore, the limit $%
Sr_{S}/V\rightarrow 0$\ can now be applied. \ This limit means that $r_{S}$\
is very small compared to the size of the averaging volume $V/S$, but there
is no requirement that $V/S$\ be a length scale at which energy is
generated. \ Thus, the size of the averaging volume is arbitrary provided
that $V/S\gg r_{S}$. \ The limit $Sr_{S}/V\rightarrow 0$\ applied to (\ref%
{TRD1}) gives

\begin{eqnarray}
\frac{r_{S}}{3}\left\langle A_{ii}\right\rangle _{\mathbb{R}} &=&\frac{r_{S}%
}{3}\partial _{\text{\c{T}}}\left\langle d_{ii}\right\rangle _{\mathbb{R}}+%
\frac{r_{S}}{3}\left\langle \left\langle \breve{u}_{n}\right\rangle _{%
\mathbb{R}}\partial _{X_{n}}d_{ii}\right\rangle _{\mathbb{R}}  \nonumber \\
+\frac{r_{S}}{3}\partial _{r_{n}}\left\langle d_{iin}\right\rangle _{\mathbb{%
R}} &=&-\frac{4}{3}r_{S}\left\langle \frac{\varepsilon +\varepsilon ^{\prime
}}{2}\right\rangle _{\mathbb{R}}.  \label{TRD22}
\end{eqnarray}%
The two terms that explicitly describe effects of inhomogeneity are
eliminated from (\ref{TRD1}). \ Of course, (\ref{TRD22}) contains the
following advective derivative%
\begin{equation}
\frac{r_{S}}{3}\left\langle \left\langle \breve{u}_{n}\right\rangle _{%
\mathbb{R}}\partial _{X_{n}}d_{ii}\right\rangle _{\mathbb{R}}=\frac{1}{3}%
\frac{Sr_{S}}{V}\oint_{\mathbf{X}_{n}}\left\langle \breve{u}%
_{n}\right\rangle _{\mathbb{R}}d_{ii}  \label{advective relation}
\end{equation}%
which is seen to vanish in the limit $Sr_{S}/V\rightarrow 0$. \ The reason
for not eliminating this advective term from (\ref{TRD22}) in the previous
step is to make a point about random sweeping at the end of this section.

\qquad The theorem of Duchon and Robert\cite{DuchonRobert00} applies to the
case $t=t^{\prime }$. \ In the previous section where the average\ was over
the entire flow we had $\left\langle \varepsilon \right\rangle _{\mathbb{R}%
}=\left\langle \varepsilon ^{\prime }\right\rangle _{\mathbb{R}}$\ if\textbf{%
\ }$t=t^{\prime }$. \ That is not true here because $\mathbb{R}$\ is now a
subdomain of the entire flow; as $\mathbf{X}$\ varies over $\mathbb{R}$\
with $\mathbf{r}$\ fixed, not every spatial point occupied by $\mathbf{x}$\
coincides with\ a spatial point occupied by $\mathbf{x}^{\prime }$ and vice
versa. \ However, including the $\mathbf{r}$-sphere\ average does cause $%
\mathbf{x}$\ and $\mathbf{x}^{\prime }$\ to equally occupy every\ spatial
point in the double integration. \ That is, for $t=t^{\prime }$, $%
\left\langle \left\langle \varepsilon \right\rangle _{\mathbb{R}%
}\right\rangle _{\mathbf{r}\text{-sphere}}=\left\langle \left\langle
\varepsilon ^{\prime }\right\rangle _{\mathbb{R}}\right\rangle _{\mathbf{r}%
\text{-sphere}}$. \ Henceforth, consider only the case $t=t^{\prime }$. \
Thus,%
\[
\left\langle \left\langle \frac{\varepsilon +\varepsilon ^{\prime }}{2}%
\right\rangle _{\mathbb{R}}\right\rangle _{\mathbf{r}\text{-sphere}%
}=\left\langle \left\langle \varepsilon \right\rangle _{\mathbb{R}%
}\right\rangle _{\mathbf{r}\text{-sphere}}=\left\langle \left\langle
\varepsilon ^{\prime }\right\rangle _{\mathbb{R}}\right\rangle _{\mathbf{r}%
\text{-sphere}}.
\]%
Apply the $r$-sphere\ average (\ref{r sphere average}) and the divergence
theorem (\ref{divergence theorem r space}), and neglect the advective term
in (\ref{advective relation}), and apply the time average, then (\ref{TRD22}%
) gives%
\begin{equation}
\frac{r_{S}}{3}\left\langle \partial _{\text{\c{T}}}\left\langle
\left\langle d_{ii}\right\rangle _{\mathbb{R}}\right\rangle _{\mathbf{r}%
\text{-sphere}}\right\rangle _{T}+\left\langle \oint_{\mathbf{r}%
_{n}}\left\langle d_{iin}\right\rangle _{\mathbb{R}}\right\rangle _{T}=-%
\frac{4}{3}r\left\langle \left\langle \left\langle \varepsilon \right\rangle
_{\mathbb{R}}\right\rangle _{\mathbf{r}\text{-sphere}}\right\rangle _{T}.
\label{TDR1}
\end{equation}%
From (\ref{ave time deriv}), the time-derivative term in (\ref{TDR1})\ is%
\begin{eqnarray}
&&\frac{r_{S}}{3}\left\langle \partial _{\text{\c{T}}}\left\langle
\left\langle d_{ii}\right\rangle _{\mathbb{R}}\right\rangle _{\mathbf{r}%
\text{-sphere}}\right\rangle _{T}  \nonumber \\
&=&\frac{r_{S}}{3T}\left[ \left\langle \left\langle d_{ii}\left( \mathbf{X,r}%
,\text{\c{T}}_{0}+T,0\right) \right\rangle _{\mathbb{R}}\right\rangle _{%
\mathbf{r}\text{-sphere}}-\left\langle \left\langle d_{ii}\left( \mathbf{X,r}%
,\text{\c{T}}_{0},0\right) \right\rangle _{\mathbb{R}}\right\rangle _{%
\mathbf{r}\text{-sphere}}\right]  \label{TavedTderiv}
\end{eqnarray}%
If $\left\langle \left\langle d_{ii}\right\rangle _{\mathbb{R}}\right\rangle
_{\mathbf{r}\text{-sphere}}$\ decreases as $r_{S}$\ decreases, then there is
an averaging duration $T$\ sufficiently large that the left-most term in (%
\ref{TDR1}) may be neglected. \ The classic inertial-range case serves as an
example; namely, $\left\langle \left\langle d_{ii}\right\rangle _{\mathbb{R}%
}\right\rangle _{\mathbf{r}\text{-sphere}}$\ scales with $\epsilon
^{2/3}r_{S}^{2/3}$, where, for brevity, $\epsilon \equiv \left\langle
\left\langle \varepsilon \right\rangle _{\mathbb{R}}\right\rangle _{\mathbf{r%
}\text{-sphere}}$. For that case, the ratio of the right-hand side of (\ref%
{TavedTderiv}) to the right-hand side of (\ref{TDR1}) is not greater than
order $\epsilon ^{-1/3}r_{S}^{2/3}/T$. \ That is, to neglect the
time-derivative term the averaging duration $T$\ must be much greater than
the inertial-range time scale $\epsilon ^{-1/3}r_{S}^{2/3}$. \ Recall that $%
\nu =0$\ such that $r_{S}$, and also $\epsilon ^{-1/3}r_{S}^{2/3}$, may be
as small as desired. \ Hence $T$\ may be much smaller than integral time
scales.

\qquad Thus, with $\nu =0$ and a space average of arbitrary length scale,
provided that length scale is much greater than $r_{S}$, and a time average
of arbitrary duration, provided that duration is much greater than the
inertial-range time scale based on $r_{S}$, we have 
\begin{equation}
\left\langle \oint_{\mathbf{r}_{n}}\left\langle d_{iin}\right\rangle _{%
\mathbb{R}}\right\rangle _{T}=-\frac{4}{3}r_{S}\left\langle \left\langle
\left\langle \varepsilon \right\rangle _{\mathbb{R}}\right\rangle _{\mathbf{r%
}\text{-sphere}}\right\rangle _{T}.  \label{TDR3}
\end{equation}%
This is closely related to the theorem of Duchon and Robert\cite%
{DuchonRobert00}. \ The order of the averaging operations in (\ref{TDR3}) is
the same as in \cite{DuchonRobert00}, although those averaging operations
commute (see Sec. 3.6). \ Note that the orientation average,\ as defined in (%
\ref{orientation average}), appears in (\ref{TDR3}) as a consequence of the
divergence theorem (\ref{divergence theorem r space}).

\subsection{Random sweeping}

\qquad Eyink\cite{Eyink03} inquires into the feasibility of numerical and
experimental tests of his theorems and the theorem of Duchon and Robert. \
An aspect of such testing is the fact that the sum of the time derivative
and advective terms in (\ref{TRD22})\ can be much smaller than either term
taken separately. The time derivative and advective terms in (\ref{TRD22})
constitute a comoving derivative. \ For simplicity, remove the factor $%
r_{S}/3$\ and the $\mathbf{r}$-sphere\ average from these two terms; then
their sum can be written as%
\begin{eqnarray*}
\partial _{\text{\c{T}}}\left\langle d_{ii}\right\rangle _{\mathbb{R}%
}+\left\langle \left\langle \breve{u}_{n}\right\rangle _{\mathbb{R}}\partial
_{X_{n}}d_{ii}\right\rangle _{\mathbb{R}} &=&\partial _{\text{\c{T}}%
}\left\langle d_{ii}\right\rangle _{\mathbb{R}}+\left\langle \breve{u}%
_{n}\right\rangle _{\mathbb{R}}\left\langle \partial
_{X_{n}}d_{ii}\right\rangle _{\mathbb{R}} \\
&=&\partial _{\text{\c{T}}}\left\langle d_{ii}\right\rangle _{\mathbb{R}}+%
\frac{S}{V}\left\langle \breve{u}_{n}\right\rangle _{\mathbb{R}}\left( \frac{%
1}{S}\int \int \widehat{N}_{n}d_{ii}dS\right) ,
\end{eqnarray*}%
wherein two of many ways to express the sum are given. \ The last expression
contains the product of the scalar $d_{ii}$\ and the outward normal unit
vector $\widehat{N}_{n}$\ averaged over the surface of the averaging volume
followed by the inner product with $\left\langle \breve{u}_{n}\right\rangle
_{\mathbb{R}}$. \ Recall that $\breve{u}_{n}\equiv \left(
u_{n}+u_{n}^{\prime }\right) /2$; thus $\left\langle \breve{u}%
_{n}\right\rangle _{\mathbb{R}}$ constitutes the sweeping velocity of the
scales larger than the averaging volume as well as any mean flow advection.
\ The effect of changes of the advective term is a corresponding change of
the time derivative $\partial _{\text{\c{T}}}\left\langle
d_{ii}\right\rangle _{\mathbb{R}}$\ such that the sum of the two terms can
be much smaller than either taken separately.

\subsection{Acceleration-velocity structure function}

\qquad The relationship of the acceleration-velocity structure function to
the energy dissipation rate is given in (\ref{MannOttIR}) and (\ref%
{MannOttsecond}) on the basis of the ensemble average without and with the $%
r $-sphere average, respectively, and in (\ref{NieTanveer accelvelo}) on the
basis of an average over the entire flow. Now, (\ref{TRD22}) contains%
\begin{equation}
\left\langle \left( a_{i}-a_{i}^{\prime }\right) \left( u_{i}-u_{i}^{\prime
}\right) \right\rangle _{\mathbb{R}}=-\left\langle \varepsilon +\varepsilon
^{\prime }\right\rangle _{\mathbb{R}}.  \label{accel velo}
\end{equation}%
This is obtained on the basis of $\nu =0$\ and the local space average
without the time and $\mathbf{r}$-sphere\ averages and without requiring $%
t=t^{\prime }$. Terms describing random and mean-flow sweeping are absent
from the derivation of (\ref{accel velo}).

\section{Effect of large-scale inhomogeneities at small scales}

\qquad Recent research has quantified the influence of inhomogeneous terms
on the balance of structure function equations. \ Those studies have applied
to channel flow\cite{Danaila01,Danaila02,Danaila04}, wind tunnel
grid-generated turbulence\cite{Lindborg99,Zhouetal00,Danaila02,Danaila04},
and turbulent jets\cite{DanailaNJP04}. \ The Reynolds decomposition of all
terms in the exact equation (\ref{basic ij eqn}) is given in \cite{Hilllanl}%
. \ That decomposition contains all possible terms contributing to the
effects of inhomogeneous large-scale structures. \ For the ensemble average,
the Reynolds decomposition of $u_{i}(\mathbf{x},t)$ is defined by 
\[
u_{i}(\mathbf{x},t)\equiv U_{i}(\mathbf{x},t)+\widehat{u}_{i}(\mathbf{x},t),%
\text{ }U_{i}(\mathbf{x},t)\equiv \left\langle u_{i}(\mathbf{x}%
,t)\right\rangle _{E},\text{ }\left\langle \widehat{u}_{i}(\mathbf{x}%
,t)\right\rangle _{E}=0,
\]%
and similarly at the point $\mathbf{x}^{\prime }$. \ Consider the Reynolds
decomposition of the average of the term $\partial _{X_{n}}F_{ijn}$ in (\ref%
{basic ij eqn})

\begin{equation}
\partial _{X_{n}}\left\langle \digamma _{ijn}\right\rangle _{E}=\frac{%
U_{n}+U_{n}^{\prime }}{2}\partial _{X_{n}}D_{ij}+\partial _{X_{n}}\left(
\Delta _{i}\widehat{\Gamma }_{jn}+\Delta _{j}\widehat{\Gamma }_{in}+\widehat{%
\Gamma }_{ijn}\right) ,  \label{advective term}
\end{equation}%
where, for brevity, we define\ 
\begin{eqnarray*}
D_{ij} &\equiv &\left\langle d_{ij}\right\rangle _{E}\text{, }\Delta
_{i}\equiv \left( U_{i}-U_{i}^{\prime }\right) , \\
\widehat{\Gamma }_{in} &\equiv &\left\langle \left( \widehat{u}_{i}-\widehat{%
u}_{i}^{\prime }\right) \frac{\widehat{u}_{n}+\widehat{u}_{n}^{\prime }}{2}%
\right\rangle _{E}, \\
\widehat{\Gamma }_{ijn} &\equiv &\left\langle \left( \widehat{u}_{i}-%
\widehat{u}_{i}^{\prime }\right) \left( \widehat{u}_{j}-\widehat{u}%
_{j}^{\prime }\right) \frac{\widehat{u}_{n}+\widehat{u}_{n}^{\prime }}{2}%
\right\rangle _{E}.
\end{eqnarray*}%
Note that $D_{ij}$\ is the structure function of velocity, not of velocity
fluctuations. \ For grid-generated turbulence data, one can let subscript 1
denote the downstream direction and approximate $U_{1}=U_{1}^{\prime }$,
such that the first term on the right-hand side of (\ref{advective term}) is 
$U_{1}\partial _{X_{1}}D_{ij}$; this is the form of the inhomogeneous term
derived in the above-mentioned studies.\cite%
{Lindborg99,Zhouetal00,Danaila02,Danaila04} \ One can easily show that the
second term on the right-hand side of (\ref{advective term}) is a
generalization of the inhomogeneous term given in \cite{Danaila01,Danaila02}
for the case of uniform channel flow. \ Of course, (\ref{advective term})
and the other Reynolds decompositions given\ in \cite{Hilllanl} are exact
for every flow, whether laminar or turbulent.\ \ Equations for structure
functions of velocity fluctuations differ significantly from equations for
the velocity; that topic has been treated in detail in \cite{Hilllanl}.

\qquad The exact structure functions are useful in other respects. \ For
example, from (\ref{taoij}) and (\ref{tao trace}) it is clear that the
pressure velocity term vanishes from the trace equations such as (\ref%
{MannOtt first}) and (\ref{space ave trace}) on the basis of
incompressibility and local homogeneity. \ It is not necessary to assume the
more restrictive basis of local isotropy as in \cite{DanailaNJP04}. \ The
combination of (\ref{basic ij eqn}) and (\ref{advective term}) shows that
both $\partial _{\text{\c{T}}}D_{ij}$\ and $\frac{1}{2}\left(
U_{n}+U_{n}^{\prime }\right) \partial _{X_{n}}D_{ij}$\ must appear in the
structure function equation as was correctly deduced by Lindborg \cite%
{Lindborg99} on the basis of mean-flow Galilean invariance, but replacing $%
\partial _{\text{\c{T}}}D_{ij}$ with $\frac{1}{2}\left( U_{n}+U_{n}^{\prime
}\right) \partial _{X_{n}}D_{ij}$, as was done in early derivations on the
basis of Taylor's hypothesis, does not preserve that invariance.

\section{Summary}

\qquad The formulation of the exact statistical equations in the variables $%
\mathbf{X}$ and\ $\mathbf{r}$ of (\ref{variable change}) produces the
pragmatic definition of local homogeneity discussed in Sec. 2. \ It is the
single definition that simplifies structure function equations for the case
of local homogeneity. \ The analogous definition of local stationarity
arises from use of the temporal variables\ \c{T} and\ \c{t} of (\ref%
{variable change}). \ The resultant organization of the exact statistical
equations provides immediate and improved derivation of the inertial-range
relationships (\ref{MannOtt first})--(\ref{MannOttsecond}) between the
acceleration-velocity structure function, the third-order velocity structure
function, and the energy dissipation rate. \ Further, all terms that must be
evaluated to include the effects of turbulence inhomogeneity and anisotropy
are evident in the formulation. \ That makes the exact statistical equation
method useful to particle tracking experiments such as the turbulence
acceleration experiments at Cornell University. \ The recent theorems of Nie
and Tanveer \cite{NieTanveer99} in (\ref{Duychon and Rob}) and of Duchon and
Robert \cite{DuchonRobert00} in (\ref{TDR3}) are likewise obtained easily,
as are their generalizations, e.g., (\ref{space ave trace}). \ The
relationship between the acceleration-velocity structure function and energy
dissipation rate is obtained in (\ref{accel velo}) on the basis of only a
local spatial average for $\nu =0$. \ With (\ref{advective term}) as one
example of the Reynolds decomposition of all terms in the exact statistical
equations in \cite{Hilllanl}, it is evident that the exact statistical
equation formulation efficiently reveals all terms that must be evaluated to
quantify the influence of inhomogeneity on the balance of structure function
equations. \ It is not necessary to derive the individual terms that
describe the effects of inhomogeneity that are missing from equations valid
only for homogeneous turbulence. \ All such terms are now known.

\qquad The above are examples of the use of exact statistical equations. \
Their usefulness arises from the organization of the equations.

\end{document}